\documentclass[aps,prb,10pt,twocolumn,superscriptaddress,floatfix]{revtex4-2}

\usepackage[utf8]{inputenc}
\usepackage{tabularx}
\usepackage{array}
\usepackage{float}
\usepackage{graphicx}
\usepackage{longtable}
\usepackage{amsmath}
\usepackage{amssymb}
\usepackage{textcomp}
\usepackage[dvipsnames]{xcolor}
\usepackage{gensymb}
\usepackage{titlesec}

\definecolor{map}{rgb}{108, 0, 108}

\definecolor{pink}{rgb}{1, 0.5, 0}

\definecolor{darkgreen}{rgb}{0, 0.5, 0}

\begin{document}

\title{Divalent Yb in the Ca$_3$Pd$_4$Bi$_8$ Structure Type: Crystal Growth and Physical Properties of Yb$_3$Pd$_4$Bi$_8$}
\author{Midori Amano Patino}
\email{midori.amano-patino@neel.cnrs.fr}
\affiliation{Univ.~Grenoble Alpes, CNRS, Grenoble-INP, Institut N\'eel, 38000 Grenoble, France}
\affiliation{Univ.~Grenoble Alpes, CEA, Grenoble-INP, IRIG, Pheliqs, 38000 Grenoble, France}
\author{Sitaram Ramakrishnan}
\affiliation{Univ.~Grenoble Alpes, CNRS, Grenoble-INP, Institut N\'eel, 38000 Grenoble, France}
\author{Davide Donadio}
\affiliation{Department of Chemistry, University of California, Davis, California 95616, USA}
\author{Nils Lamouche}
\affiliation{Département de Physique, Université de Montréal, Montréal, Quebec H2V 0B3, Canada}
\author{Elise Pachoud}
\affiliation{Univ.~Grenoble Alpes, CNRS, Grenoble-INP, Institut N\'eel, 38000 Grenoble, France}
\author{Macha Meplan}
\affiliation{Univ.~Grenoble Alpes, CEA, Grenoble-INP, IRIG, Pheliqs, 38000 Grenoble, France}
\author{Daniel Braithwaite}
\affiliation{Univ.~Grenoble Alpes, CEA, Grenoble-INP, IRIG, Pheliqs, 38000 Grenoble, France}
\author{Jean-Pascal Brison}
\affiliation{Univ.~Grenoble Alpes, CEA, Grenoble-INP, IRIG, Pheliqs, 38000 Grenoble, France}
\author{Gerard Lapertot}
\affiliation{Univ.~Grenoble Alpes, CEA, Grenoble-INP, IRIG, Pheliqs, 38000 Grenoble, France}
\author{Andrea Bianchi}
\affiliation{Département de Physique et Regroupement Qu\'eb\'ecois sur les Mat\'eriaux de Pointe (RQMP), Université de Montréal, Montréal, Canada, and Institut Courtois Université de Montréal, Montréal, Quebec H2V 0B3, Canada}
\author{Georg Knebel}
\affiliation{Univ.~Grenoble Alpes, CEA, Grenoble-INP, IRIG, Pheliqs, 38000 Grenoble, France}
\author{Valentin Taufour}
\affiliation{Department of Physics and Astronomy, University of California, Davis, California 95616, USA}

\begin{abstract}
We report on the single crystal growth, crystal structure, and physical properties of Yb$_3$Pd$_4$Bi$_8$. Single crystal x-ray diffraction shows that the material crystallizes in the Ca$_3$Pd$_4$Bi$_8$ structure-type at ambient conditions, described by orthorhombic symmetry $Pbam$. Magnetization measurements reveal a nearly isotropic diamagnetic response, indicating that Yb is stabilized in a non-magnetic divalent configuration. Electrical resistivity measurements show a metallic behavior. First-principles
calculations including spin--orbit coupling describe a metal whose states at the
Fermi level are dominated by Bi $6p$ orbitals, with a density of states within
1\% of that of the isoelectronic compound Ca$_3$Pd$_4$Bi$_8$, consistent with a
filled and chemically inert $4f$ shell. They further show that the
nonsymmorphic symmetry of the crystal enforces fourfold-degenerate nodal lines on the
Brillouin-zone boundary, four of which cross the Fermi level and give
symmetry-enforced Dirac points with a linear transverse dispersion. Despite strong metallic and covalent bonding within the Pd–Bi network, Yb$_3$Pd$_4$Bi$_8$ is realized as a rare-earth analog of the Ca$_3$Pd$_4$Bi$_8$ structure type with divalent Yb ions. This result offers a potential avenue to discover other rare-earth analogs within this structural family.
\end{abstract}

\maketitle

\section{Introduction}

In condensed matter physics and materials science, the drive to synthesize entirely new materials with novel crystal structures is the primary engine for discovering unprecedented quantum phenomena. 
New Yb-based intermetallic compounds have played an important role in understanding strongly correlated electron physics. This is mainly due to the fact that the valence of the Yb ion can fluctuate between a non-magnetic $\text{Yb}^{2+}$ ($4f^{14}$) configuration and a magnetic $\text{Yb}^{3+}$ ($4f^{13}$) configuration \cite{Varma1976RMP}. Compounds with either of the two configurations exist depending on the local environment of the Yb and the hybridization strength with the conduction electrons. Several Yb-based heavy fermion systems have been at the forefront of research. For example, YbRh$_2$Si$_2$ is a prime example for heavy fermion quantum criticality \cite{Trovarelli2000, Knebel2006JPSJ, Gegenwart2008NatPhys}; half-Heusler compounds like $\text{YPtBi}$ are famous for hosting topological semimetal and superconductivity with a very high angular momentum Cooper pairing \cite{Chadov2010NM, Guo2018, Kim2018sciadv}; YbInCu$_4$ shows a spectacular first order valence transition from the Yb$^{3+}$ to an intermediate Yb valence as a function of temperature \cite{Sarrao1998}; or YbB$_{12}$ is a prime candidate for a topological Kondo Insulator \cite{Kasaya1983, Kasuya1994}. Despite having an insulating bulk gap driven by $4f$ correlations, quantum oscillations have been observed at high fields, pointing to exotic, topologically protected metallic surface states \cite{Hagiwara2016, Xiang2018a, Mizzi2024, Yang2024}

Recently, Pd-Bi compounds have been intensively studied, as they are hallmark systems for physics with strong spin-orbit coupling and superconductivity. The extremely strong spin-orbit coupling in Bi often leads to topological band properties with band inversion and topological surface states. Pd-Bi systems like $\text{PdBi}$ and $\beta\text{-PdBi}_2$ are intrinsic bulk superconductors with topological states on the surface. \cite{Imai2012JPSJ, Sakano2015,  Powell2025}. This means that the topological surface states and the superconducting gap exist in the same system. Introducing Yb into the Pd - Bi system can add strong electronic correlations and open the way to heavy-fermion physics. As an example, YbPdBi is a half-Heusler heavy fermion compound with a strongly enhanced Sommerfeld coefficient of the specific heat ($\gamma \approx 1.2\text{ J/mol}\cdot\text{K}^2$ with a possible antiferromagnetic order near 1\,K \cite{LeBras1995JPCM, Kaczorowski1999PRB, Pietri2000PRB}. Due to its non-centrosymmetric crystal structure (space group $F\bar{4}3m$) it is a candidate Weyl semimetal, similar to YbPtBi. 

Beyond the known half-Heusler phase, the Yb-Pd-Bi system also offers an opportunity for discovering new rare-earth intermetallics. In this respect, the reported alkaline-earth compound $\mathrm{Ca_3Pd_4Bi_8}$~\cite{Johrendt2002ZAAC,Ovchinnikov2022IC} is a particularly useful structural prototype. Its Pd-Bi framework accommodates divalent alkaline-earth cations, suggesting that substitution by Yb may be possible if ytterbium adopts its divalent, non-magnetic $4f^{14}$ configuration. The $\mathrm{Ca_3Pd_4Bi_8}$ crystal structure, $Pbam$, is centrosymmetric, so that in a non-magnetic compound the combination of inversion and time reversal makes every band 2-fold degenerate throughout the zone and forbids Weyl nodes. At the same time, it is nonsymmorphic: its two glide planes carry half-lattice translations, and force bands to be 4-fold degenerate on parts of the zone boundary~\cite{Young2012PRL}. Together with the strong spin--orbit coupling of Bi, this makes the Ca$_3$Pd$_4$Bi$_8$ framework a natural setting for symmetry-enforced Dirac degeneracies, a question that the two existing scalar-relativistic studies of Ca$_3$Pd$_4$Bi$_8$ could not address exhaustively~\cite{Johrendt2002ZAAC,Ovchinnikov2022IC}.

Here we report on the synthesis, crystal structure and physical properties of single-crystalline Yb$_3$Pd$_4$Bi$_8$. It crystallizes in the Ca$_3$Pd$_4$Bi$_8$ structure type. Magnetization,
transport and heat-capacity measurements indicate metallic behavior with
divalent Yb. These measurements are complemented by fully
relativistic density-functional calculations, which place the Fermi level in a
Bi $6p$ band complex essentially identical to that of
Ca$_3$Pd$_4$Bi$_8$, and which locate four symmetry-enforced Dirac points at the
Fermi level on the 4-fold nodal lines of the zone boundary.

\section{Methods}

\subsection{Crystal growth and experimental probes}

Single crystals of Yb$_3$Pd$_4$Bi$_8$ were grown using a self-flux method. The clean metals, Yb (\( \geq 99\% \), G. Holdron), Pd (Alfa Aesar, 99.98+\%) and Bi (Puratronic(R), 99.98+\%), were introduced in a Canfield crucible set~\cite{Canfield2016PM,Canfield2025ZAAC} in the 1:1:7, Yb:Pd:Bi molar ratio and sealed in an evacuated silica ampoule. The ampoule was heated to 1150\,\degree C in 48\,h, held there for 8\,h, and slowly cooled to 500\,\degree C at 1.6\,\degree C per hour. At 500\,\degree C, the formed crystals were filtered by removing the ampoule from the furnace, inverting it, and immediately spinning it in a centrifuge~\cite{Paglione2021}. 
The as-synthesized Yb$_3$Pd$_4$Bi$_8$ single crystals exhibit a metallic gray color and needle-like morphology.

Energy dispersive X-ray analysis performed in a FESEM Zeiss Ultra+ scanning electron microscope with a Bruker SDD detector confirmed the Yb$_3$Pd$_4$Bi$_8$ stoichiometry. 

Single-crystal X-ray diffraction data were acquired at the side-station of ID28 beamline at European Synchrotron Radiation Facility (ESRF), Grenoble, France, employing a wavelength ($\lambda$) of 0.5701 \AA. The detector used is a Dectris PILATUS3 1M area detector (pixel size 172 $\times$ 172 $\mu$m) on a four-circle Euler diffractometer with angular step rotation of 0.25$^\circ$ and exposure of 0.25 seconds. Further crystallographic information can be inferred from Table \ref{tab:Ybp_crystalinfo}. Integration and subsequent refinement of the crystal structure was done using CRYSALISPRO \cite{crysalis} and JANA 2006 \cite{petricekv2014a}.

\begin{table}[ht]
\caption{\label{tab:Ybp_crystalinfo}%
Crystallographic data of Yb$_3$Pd$_4$Bi$_8$ at 293 K. The criterion of observability is $I>3\sigma(I)$.}
\centering
\begin{ruledtabular}
\begin{tabular}{cc}
Crystal system & Orthorhombic  \\
Space group & $Pbam$    \\
Space group no. & 55 \\
$a$ (\AA{}) &10.8109(4)         \\
$b$ (\AA{}) &17.0493(7)                   \\
$c$ (\AA{}) &4.1428(1)                \\
Volume (\AA{}$^3$) & 763.6(3)          \\
$Z$ & 2   \\
Wavelength (\AA{}) & 0.5701     \\
Detector distance (mm) &170   \\
Rotation per image ($\phi$) ($^\circ$)  & 0.25   \\
$\chi$ offset ($^\circ$) & -45 \\
2$\theta$ offset ($^\circ$) & -19 \\
$(\sin(\theta)/\lambda)_{max}$ (\AA{}$^{-1}$) &0.611578  \\
Absorption, $\mu$ (mm$^{-1}$) & 64.806     \\
No. of reflections measured  &  1816  \\
No. of unique reflections (obs/all)& 741/741  \\
$R_{int}$ (obs/all) &0.0250/0.0250    \\
No. of parameters &47   \\
$R_{F }$ (obs/all) &0.0448/0.0448 \\
$wR_{F }$ (all)  &0.0645   \\
GoF (obs/all) &5.09/5.09  \\
$\Delta\rho_{min}$, $\Delta\rho_{max}$(e \AA$^{-3}$) &
-5.5, 4.36  \\
\end{tabular}
\end{ruledtabular}
\end{table}

\begin{table*}[ht]
\scriptsize
\caption{\label{stab:Ypb_atomic}%
Room temperature atomic coordinates $x$, $y$, $z$ and
atomic displacement parameters (ADPs) $U_{ij}$
of Yb$_3$Pd$_4$Bi$_8$ with space group $Pbam$.
During refinement, the $U_{22}$ parameter of the Bi1 site was constrained to a small positive value. When refined freely, this parameter became negative, resulting in a non-positive definite ADP tensor. This behavior is most likely attributable to minor imperfections in the absorption correction, as the compound is a strong X-ray absorber, and even small inaccuracies in absorption correction can disproportionately affect anisotropic displacement parameters. However, this constraint has a negligible effect on the overall structural model.}
\begin{ruledtabular}
\begin{tabular}{lcccccccccc}
Atom & $x$ & $y$ & $z$ & $U_{11}$ & $U_{22}$ & $U_{33}$ & $U_{12}$ & $U_{13}$ & $U_{23}$ & $U^{eq}_{iso}$ \\
\hline
\multicolumn{11}{l}{Orthorhombic structure of Yb$_3$Pd$_4$Bi$_8$ at T = 293 K, SG: $Pbam$}  \\
\hline
Yb1  & 0 & 0 & 0 &  0.039(1) &  0.00042(9) & 0.0269(9)& -0.0089(8) & 0 & 0 & 0.0235(6) \\
Yb2 &0.3572(1)  & 0.19978(8) &  0 & 0.0202(8)&  0.0051(7) & 0.0233(7)& -0.0048(5) & 0 & 0 & 0.0162(4) \\
Pd1 &0.4540(2)  & 0.0788(1) &  0.5 & 0.0177(14) & 0.0014(12) &  0.0306(12) & 0.0005(9) & 0 & 0 & 0.0166(7) \\
Pd2 & 0.0718(2) & 0.17623(12) & 0 & 0.0229(14) & 0.0036(12) & 0.020(10) & 0.0017(10) & 0 & 0 & 0.0156(7) \\
Bi1 & 0.1100(1) & 0.28681(6) & 0.5 & 0.0172(7) & 0.0001 & 0.0227(6) & 0.0012(4) & 0 & 0 & 0.0133(3) \\
Bi2 & 0.1958(1) & 0.08565(6) & 0.5 & 0.0148(7) & 0.0030(6) & 0.0230(6) & 0.0019(4) & 0 & 0 & 0.0136(4) \\
Bi3 & 0.3907(1) & 0.35729(6) & 0.5 & 0.0162(7) & 0.0043(6) & 0.0254(7) & -0.0004(4) & 0 & 0 & 0.0153(4) \\
\end{tabular}
\end{ruledtabular}
\end{table*}


Magnetization measurements were performed using an MPMS3 SQUID magnetometer (Quantum Design) in VSM scan mode down to 2\,K and up to 7\,T.

Resistivity measurements were performed in a physical property measurement system (PPMS, Quantum Design) using a four-point lock-in technique; 15\,$\mu$m Au wires were spot-welded to the sample for electrical contacts. 
The specific heat was measured using the PPMS with the heat pulse relaxation method. 

\subsection{First-principles calculations}

Electronic-structure calculations were performed with the plane-wave
pseudopotential code \textsc{Quantum ESPRESSO} (v7.2)~\cite{Giannozzi2017JPCM}
using the PBE generalized-gradient functional~\cite{Perdew1996PRL} and
projector augmented-wave pseudopotentials~\cite{Blochl1994PRB} from
\textsc{PSLibrary} 1.0.0~\cite{DalCorso2014CMS}. For Yb the
$4f$ shell is treated as core states, so that Yb is divalent by construction, as established by the diamagnetic susceptibility measurements. 
This choice makes Yb and Ca isovalent, $z=10$, with 444 valence electrons in the
30-atom cell. Wavefunction and charge-density cutoffs of 70 and 700\,Ry were
used, well above the values suggested by the dataset generation tests (56 and
539\,Ry for Pd, the most demanding species). These values were chosen by running convergence tests on total energies for cutoffs between 40 and 90~Ry.
Spin--orbit coupling was included non-perturbatively through fully relativistic pseudopotentials in a noncollinear
calculation; the scalar-relativistic results are reported alongside for comparison.

Ionic positions were relaxed at the experimental 293\,K cell with a
$4\times3\times10$ shifted Monkhorst--Pack mesh~\cite{Monkhorst1976PRB} and
Marzari--Vanderbilt smearing~\cite{Marzari1999PRL}, until residual forces fell
below $10^{-3}$\,Ry/bohr. Densities of states were obtained from a
$12\times8\times32$ shifted mesh, 384 points in the irreducible wedge, using
the optimized tetrahedron method~\cite{Kawamura2014PRB}; the integrated density
of states recovers 444.0 electrons in every calculation reported. Orbital
characters were obtained by projection onto L\"owdin-orthogonalized atomic
orbitals~\cite{Lowdin1950JCP}, which recover 98\,\% of the density of states at the Fermi level $N(E_\mathrm{F})$. Band
structures were computed along
$\Gamma$--X--S--Y--$\Gamma$--Z--U--R--T--Z, Y--T, U--X and S--R with 348
$k$ points. Two further cuts of 41 $k$ points each, one along
$a^*$ and one along $b^*$ through the point $(\tfrac{1}{2},\tfrac{1}{2},0.123)$
on the S--R line, were computed to resolve the dispersion transverse to that
line.

The isostructural, isoelectronic analogue Ca$_3$Pd$_4$Bi$_8$ was computed under
an identical protocol but using only scalar relativistic pseudopotentials at its own experimental cell, which differs in volume by
only 0.18\,\%, to isolate the role of the Yb $4f$ shell. The sensitivity of the
relaxed geometry to the exchange--correlation functional was tested by
repeating the relaxation with PBEsol~\cite{Perdew2008PRL}.

\section{Results and discussion}

\subsection{Crystal structure}

\begin{figure*}[!htb]
\centering
\includegraphics[width= 150 mm]{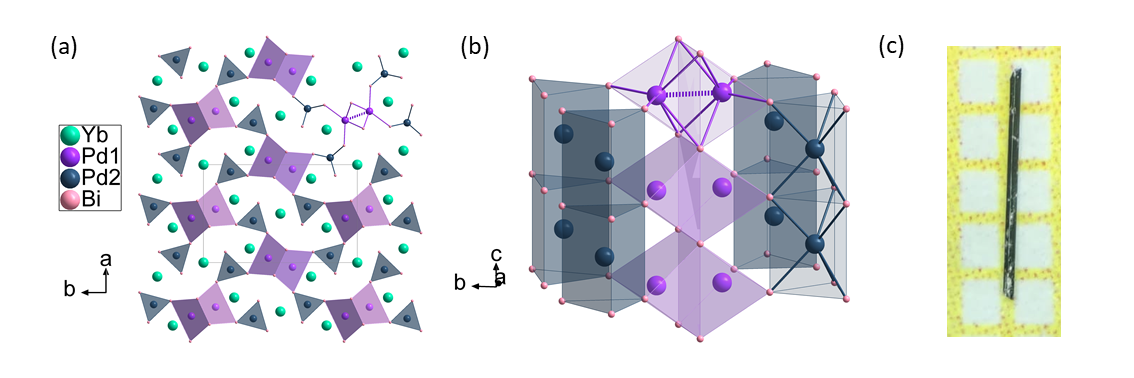}
\caption{The crystal structure of Yb$_3$Pd$_4$Bi$_8$ is shown in (a) and (b). A picture of a single crystal on a millimetre grid is shown in (c).\label{crystal_structure}} 
\end{figure*}

All diffraction maxima were indexed on a primitive orthorhombic lattice corresponding to $a$ = 10.8109(4) \AA{}, $b$ = 17.0493(7) \AA{}, $c$ = 4.1428(1) \AA{}. The structure of Yb$_3$Pd$_4$Bi$_8$ is isostructural to Ca$_3$Pd$_4$Bi$_8$~\cite{Ovchinnikov2022IC}. As depicted in Table \ref{tab:Ybp_crystalinfo}, the fit to the data employing orthorhombic $Pbam$ symmetry is excellent ($R_F$ = 4.5\%). The refined atomic coordinates and atomic displacement parameters are listed in Table \ref{stab:Ypb_atomic}. As illustrated in Figure \ref{crystal_structure}, the structure is a 3D network of Pd-Bi chains running along the $c$ axis, with Yb atoms occupying the voids. Each Pd atom is coordinated to six Bi atoms, forming distorted PdBi$_6$ trigonal prisms. Pd(1) lies at the center of Pd(1)Bi$_6$ wedged-type distorted trigonal prisms that share their rectangular bases to form Pd(1)$_2$Bi$_8$ dimeric units. These units are connected through their common edges to form the chains along the $c$ axis. The Pd-Pd distance within the dimeric unit is very short, at 2.864(5)\AA{} in the ($ab$) plane, suggesting metal-metal bonding. The second chain consists of more regular Pd(2)Bi$_6$ triangular prisms sharing their triangular faces, with a Pd-Pd distance of 4.1428(11)\AA{}. The two chains are connected by the vertices of the prisms as seen in Fig. 1b. The isostructural relationship between Yb$_3$Pd$_4$Bi$_8$ and its Ca-based analogs can be rationalized on crystal-chemical grounds. Although these compounds are far from the ionic limit and instead exhibit substantial metallic and covalent bonding contributions, the substitution of Ca by Yb is facilitated by the close match in the effective ionic radii of $\mathrm{Ca}^{2+}$ and $\mathrm{Yb}^{2+}$ \cite{Shannon1976}.  This is expected to promote the retention of the same structure type while introducing additional $4f$ electronic degrees of freedom. However, it should be emphasized that Shannon ionic radii are formulated for predominantly ionic environments with well-defined coordination geometries. Their application to complex intermetallic phases such as Yb$_3$Pd$_4$Bi$_8$ should therefore be regarded as a qualitative guide rather than a quantitative predictor of structural stability.

Nevertheless, this cation substitution approach naturally extends to rare earths for a plethora of materials \cite{Fornasini2006ZfK}. Eu-based analogs can be considered, as $\mathrm{Eu}^{2+}$ also exhibits an ionic size comparable to alkaline-earth ions such as Ca, Sr, and Ba. The transition from alkaline-earth compounds to Eu-containing counterparts are widely realized across other intermetallic families like the $X$Al4 compounds ($X$ = Ca, Sr, Ba, Eu) \cite{miller1993JACS, Parthe1983a, Ramakrishnan2024PRR, Ramakrishnan2022a} or the  $A$MgGe Zintl family of thermoelectrics ($A$ = Ca, Sr, Eu, Yb) \cite{Yuan2025JAC} and several others. Moreover, in comparison with the non-magnetic $\mathrm{Yb}^{2+}$ ($4f^{14}$) configuration, $\mathrm{Eu}^{2+}$ ($4f^{7}$) carries a large localized effective magnetic moment $\mu_{\text{eff}} \approx 7.94 \, \mu_B$. In metallic systems, these local moments are often coupled through the Ruderman--Kittel--Kasuya--Yosida (RKKY) interaction mediated by conduction electrons, giving rise to a rich variety of magnetic ground states. Thus, Eu-based compounds provide an ideal testbed for exploring the interplay between magnetism, RKKY exchange, strong spin--orbit coupling and emergent electronic phases.

\subsection{Magnetization}

\begin{figure}[!htb]
\centering
\includegraphics[width=\linewidth]{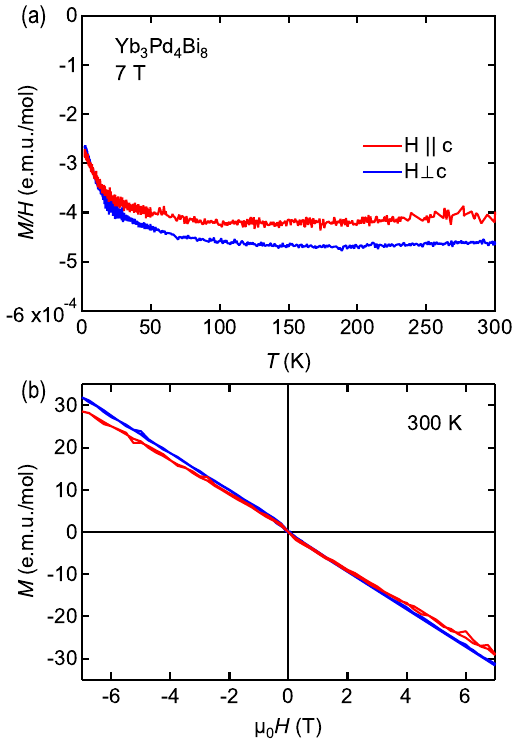}
\caption{(a) Temperature dependence of the magnetic susceptibility under an applied magnetic field of 7\,T parallel and perpendicular to the $c$ axis. (b) Magnetic field dependence of the magnetization at 300\,K for the field parallel and perpendicular to the $c$ axis.\label{Magnab}} 
\end{figure}

The magnetization data collected from a single crystal of Yb$_3$Pd$_4$Bi$_8$, with a field of 7\,T applied parallel and perpendicular to the crystallographic $c$ axis, is presented in Fig.~\ref{Magnab}(a). The magnetization isotherms $M(H)$ collected at 300\,K are shown in Fig.~\ref{Magnab}(b). The magnetic susceptibility is nearly isotropic. The negative susceptibility indicates that Yb$_3$Pd$_4$Bi$_8$ is diamagnetic, confirming that Yb is in the non-magnetic Yb$^{2+}$ configuration. The upturn at low temperature indicates a small amount of paramagnetic impurity, which is not significant enough to overcome the diamagnetic response. The diamagnetic susceptibility is around $-4\times10^{-4}$\,e.m.u./mol, which is comparable to that of Ca$_3$Pd$_4$Bi$_8$ ($\sim-2\times10^{-4}$\,e.m.u./mol) and Ca$_3$Pt$_4$Bi$_8$ ($\sim-3\times10^{-4}$\,e.m.u./mol)~\cite{Ovchinnikov2022IC}, with the larger magnitude in Yb$_3$Pd$_4$Bi$_8$ reflecting the larger number of core electrons.

\subsection{Resistivity}

\begin{figure}[!tb]
\centering
\includegraphics[width=\linewidth]{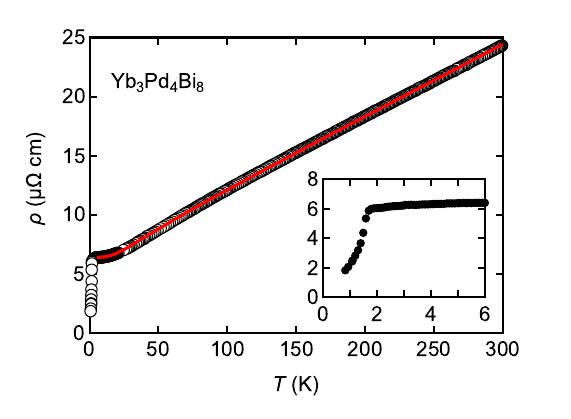}
\caption{Temperature dependence of the resistivity at zero magnetic field. The red solid line is a fit to the resistivity with the Bloch-Gr\"{u}neisen formula. The inset shows the resistivity below $6$~K. Below 1.8~K, the resistivity shows the onset of superconductivity, but its origin may be extrinsic due to Pd-Bi flux on the surface. \label{resistivity}} 
\end{figure}

Fig.~\ref{resistivity} shows the temperature dependence of the electrical resistivity of Yb$_3$Pd$_4$Bi$_8$ measured at zero magnetic field with the current applied along the crystallographic direction parallel to the needle axis, [001]. The resistivity decreases upon cooling, indicating metallic behavior. The residual resistivity ratio $RRR = \rho (300 {\rm K})/\rho (0 {\rm \,K}) \approx 4$ is rather low. The temperature dependence can be well  fitted using the Bloch-Gr\"{u}neisen formula of the resistivity,
\[
\rho(T) = \rho_0 + A \left( \frac{T}{\theta_R} \right)^5 \int_0^{\theta_R/T} \frac{x^5}{(e^x - 1)(1 - e^{-x})} \, dx
\]
Where $A$ is a prefactor and $\theta_R$ is the characteristic temperature associated with phonon-mediated electrical transport. The fit yields $\rho_0 = 6.4\,\mu\Omega$~cm, $A = 25.5\,\mu\Omega$~cm and $\theta_R \approx 105$~K. This agreement is consistent with electron–phonon scattering accounting for the temperature-dependent resistivity, without requiring an additional magnetic scattering contribution. The value of $\theta_R$ is comparable to those reported for Ca$_3$Pd$_4$Bi$_8$ and Ca$_3$Pt$_4$Bi$_8$, approximately 128 and 117~K, respectively~\cite{Ovchinnikov2022IC}. These rather low Debye temperatures may originate from soft phonon modes associated with the heavy bismuth polyanionic framework ($\text{[Pd}_4\text{Bi}_8]^{6-}$ chains). The inset in Fig.~\ref{resistivity} shows the temperature dependence below 6~K. The resistivity starts to decrease near 5.4~K and 1.8~K it decreases strongly. This may be due to the onset of superconductivity, but we cannot exclude that this superconductivity is due to some residual flux on the surface. E.g.~$\beta-\text{PdBi}_2$ (tetragonal Phase) has $T_c = 5.4$~K and $\alpha-\text{PdBi}_2$ which crystallize in a monoclinic phase and becomes superconducting below 1.7~K \cite{Imai2012JPSJ}. A magnetic field of 0.1~T is enough to completely suppress this extrinsic superconducting transition in Yb$_3$Pd$_4$Bi$_8$ (not shown). 

\subsection{Heat capacity}

\begin{figure}[!tb]
\centering
\includegraphics[width=\linewidth]{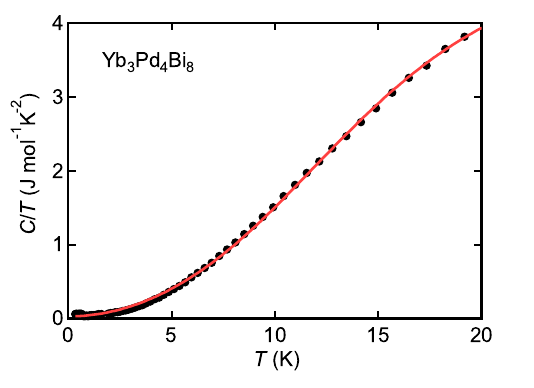}
\caption{Temperature dependence of the specific heat divided by temperature ($C/T$) at zero magnetic field below 20~K. The red line is a fit with a Debye model for the specific heat. \label{CdT}} 
\end{figure}

The temperature dependence of the specific heat divided by the temperature ($C/T$) below 20\,K is shown in Fig.~\ref{CdT}. The specific heat decreases monotonously to low temperatures without any anomalies above 1~K, supporting that the transitions seen in the resistivity are not bulk transitions. The temperature dependence of the specific heat can be fitted with the Debye model and the Debye temperature $\theta_D = 107.5$~K is in excellent agreement with $\theta_R = 105$~K from the temperature dependence of the resistivity.  At low temperature, we observed a jump of the specific heat on cooling at 0.78~K from $C/T \approx 0.035$ J mol$^{-1}$K$^{-2}$ to  $C/T \approx 0.06$ J mol$^{-1}$K$^{-2}$; however, the origin of this jump in the specific heat is still unclear.

\subsection{Electronic structure }

\begin{table}[tb]
\caption{Shortest interatomic contacts in Yb$_3$Pd$_4$Bi$_8$ ($M$~=~Yb): experiment at 293~K, and calculated values after relaxing the ionic positions at the experimental cell with the PBE and PBEsol functionals. $\Delta$ is relative to experiment. The isostructural Ca$_3$Pd$_4$Bi$_8$ is given for the Pd--Pd contact only. Differences between theory and experiments are well within thermal displacements. For example, the refined root-mean-square thermal displacement of Pd1 is 0.133\,\AA\ along $a$, against a computed displacement of 0.032\,\AA.
\label{tab:dft_bonds}}
\centering
\begin{ruledtabular}
\begin{tabular}{lccccc}
Contact & Exp.\ (\AA) & PBE (\AA) & $\Delta$ & PBEsol (\AA) & $\Delta$ \\ \hline
Pd--Pd & 2.864 & 2.923 & $+2.05$\% & 2.910 & $+1.61$\% \\
Bi--Pd & 2.794 & 2.801 & $+0.25$\% & 2.798 & $+0.12$\% \\
Bi--Bi & 3.123 & 3.178 & $+1.75$\% & 3.111 & $-0.39$\% \\
$M$--Pd & 3.103 & 3.047 & $-1.83$\% & 3.032 & $-2.31$\% \\
$M$--Bi & 3.302 & 3.270 & $-0.97$\% & 3.284 & $-0.54$\% \\ \hline
Ca$_3$Pd$_4$Bi$_8$ & & & & & \\
Pd--Pd & 2.851 & 2.859 & $+0.30$\% & --- & --- \\
\end{tabular}
\end{ruledtabular}
\end{table}

\begin{figure}[!tb]
\centering
\includegraphics[width=85mm]{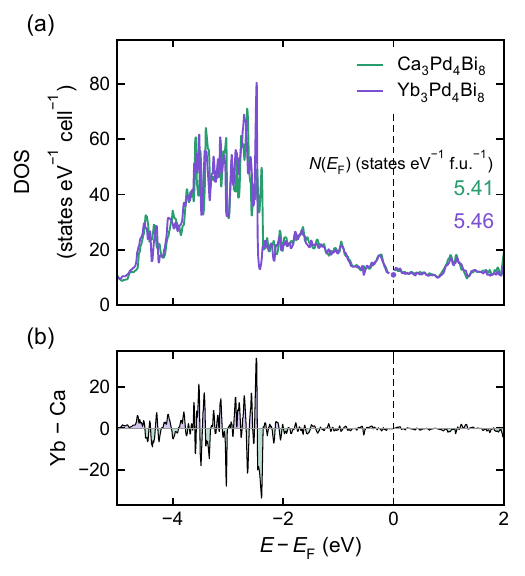}
\caption{Comparison of Yb$_3$Pd$_4$Bi$_8$ with the isoelectronic,
isostructural analogue Ca$_3$Pd$_4$Bi$_8$, computed under an identical protocol, with scalar relativistic pseudopotentials,  and aligned on $E-E_\mathrm{F}$; the two cells differ only by 0.18\% in volume.
(a) Total densities of states, per unit cell, which contains two formula units ($Z=2$). (b) Their difference: spectral weight is redistributed within the Pd $4d$ block below $-2$\,eV, while above $-2$\,eV the two curves are superimposable, with a Pearson correlation of 0.979.\label{fig:dft_spectator}}
\end{figure}

\begin{figure*}[!tb]
\centering
\includegraphics[width=160mm]{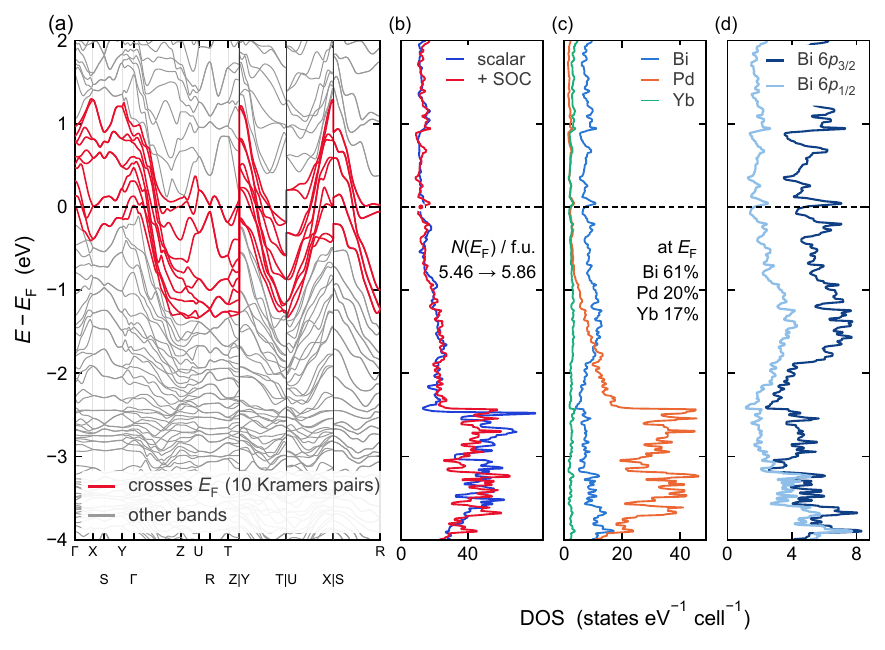}
\caption{Electronic structure of Yb$_3$Pd$_4$Bi$_8$ including spin--orbit coupling. (a) Kohn-Sham bands along the $Pbam$ high-symmetry path. Each band is occupied by an electron pair: the ten pairs that cross $E_\mathrm{F}$ are highlighted. (b) Total density of states, scalar-relativistic and with spin--orbit coupling, on the same energy axis. (c) Element-resolved density of states with spin--orbit coupling; at $E_\mathrm{F}$ the Bi, Pd and Yb contributions are 61\%, 20\% and 17\%. (d) Bi $6p$ resolved by total angular momentum. At $E_\mathrm{F}$, the $j=3/2$ to $j=1/2$ ratio is 2.9, against the value 2.0 expected from their degeneracies. The density-of-states axes in (b)--(d) are per unit cell, which contains two formula units ($Z=2$); every $N(E_\mathrm{F})$ quoted on the figure, in the text, and in Fig.~\ref{fig:dft_spectator} is per formula unit.\label{fig:dft_bands}}
\end{figure*}

Relaxing the ionic positions at the experimental 293~K cell reproduces the refined structure closely (Table~\ref{tab:dft_bonds}). The Bi--Pd contact that defines the PdBi$_6$ prisms is within 0.25\% of experiment and the cation--anion distances within 2\%. The least well reproduced contact is the short Pd--Pd bond, computed 2\% too long. 
Repeating the relaxation with PBEsol, with everything else held fixed, shortens the bond, reducing the discrepancy to $+1.6$\%, while improving Bi--Bi and leaving $M$--Pd slightly worse. 
These discrepancies, stemming from approximations in the GGA semilocal density functionals, are small in absolute terms. 
In fact, the calculated 0~K geometry agrees with the 293~K refinement to within the thermal amplitude. For example, the computed displacement of Pd1 is 0.032\,\AA\ against a refined root-mean-square thermal amplitude of 0.133\,\AA\ for that atom.

The calculations were performed with the $4f$ shell frozen in the core, so that Yb is divalent by construction, as established by the diamagnetic susceptibility reported above. 
The good agreement with the experimental geometry suggests that the Yb$^{2+}$ assignment is correct and the $4f$ electrons are spectators at the Fermi energy. 
With Yb$^{2+}$, the compound is isoelectronic with Ca$_3$Pd$_4$Bi$_8$, and both the relaxed geometry (Table~\ref{tab:dft_bonds}) and the electronic structure (Fig.~\ref{fig:dft_spectator}) of the two compounds are almost indistinguishable. Specifically, both compounds have very similar densities of states at the Fermi energy ($N(E_\mathrm{F})$) (Fig.~\ref{fig:dft_spectator}). Our $N(E_\mathrm{F})$ for Ca$_3$Pd$_4$Bi$_8$, 5.41\,states\,eV$^{-1}$ per formula unit, agrees well with the value reported in
Refs.~\onlinecite{Johrendt2002ZAAC, Ovchinnikov2022IC}.
The Pauli and Landau contributions implied by the calculated density of states, $\chi_\mathrm{Pauli}+\chi_\mathrm{Landau} = \frac{2}{3} \mu_\mathrm{B}^2 N_\mathrm{A} N(E_\mathrm{F}) = +1.3\times10^{-4}$~emu/mol, are positive and about three times smaller in magnitude than the measured magnetic susceptibility. Subtracting them from the measured susceptibility leaves a core contribution of $-5.3\times10^{-4}$~emu/mol, consistent with a response dominated by the core diamagnetism of a closed $4f^{14}$ shell.

The fully relativistic spin-orbit-coupling calculation confirms that the compound is a metal with a modest density of states at the Fermi level (Fig.~\ref{fig:dft_bands}). 
Ten bands cross $E_\mathrm{F}$, giving $N(E_\mathrm{F}) = 5.86$ states\,eV$^{-1}$ per formula unit, a modest increase over the 5.46~states\,eV$^{-1}$ per formula unit computed without spin--orbit coupling. The corresponding band value of the Sommerfeld coefficient is $\gamma_\mathrm{band} = 13.8$~mJ\,mol$^{-1}$K$^{-2}$ per formula unit. The states at the Fermi level are dominated by Bi, which contributes 61\% of $N(E_\mathrm{F})$ against 20\% for Pd and 17\% for Yb (Fig.~\ref{fig:dft_bands}c), and within the Bi contribution it is the $6p_{3/2}$ manifold that dominates. The Pd $4d$ shell is filled and lies 2--4\,eV below $E_\mathrm{F}$, so the metallic conduction is carried by the Bi--Pd framework.
The origin of the increase of $N(E_\mathrm{F})$ lies in the Bi $6p$ manifold, which spin--orbit coupling splits into $j=1/2$ and $j=3/2$ components separated by 0.66\,eV. At the Fermi level the two are populated in the ratio 2.88, whereas their degeneracies alone would give 2.00 (Fig.~\ref{fig:dft_bands}d): the $6p_{3/2}$ states are over-represented at $E_\mathrm{F}$ by 44\%. The effect is not confined to Bi, however, as the Yb, Pd and Bi contributions increase comparably.

\begin{figure}[!tb]
\centering
\includegraphics[width=85mm]{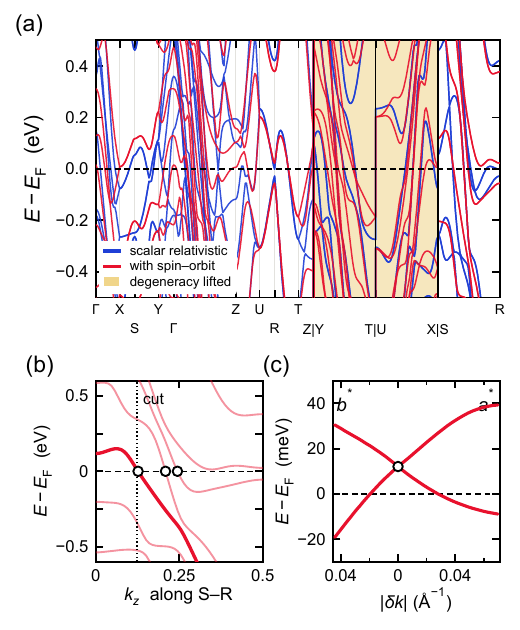}
\caption{Spin--orbit coupling and the symmetry-enforced band degeneracies. (a) Scalar-relativistic and spin--orbit band structures overlaid on the identical 348-point path, each referred to its own Fermi level. Shading marks the two segments where spin--orbit coupling destroys the degeneracy. 
The two shaded segments are the parts of the $k_x=1/2$ and $k_y=1/2$ nodal surfaces that carry no glide protection. The Kramers degeneracy is never lifted. 
(b) The fourfold branch nearest $E_\mathrm{F}$ along the S--R nodal line (heavy line); thin lines are the other fourfold branches in the window and open circles mark every crossing of $E_\mathrm{F}$. The dotted line marks where the perpendicular cuts of (c) were taken, $k_z=0.123$, which is 0.0035 reciprocal lattice units short of the exact crossing at $k_z=0.1265$. (c) The two perpendicular cuts, plotted back to back on a common momentum scale: left half along $b^*$, right half along $a^*$. Each cut is exactly even in $\delta k$, so one half of each is shown; each branch drawn is a Kramers doublet, so four states meet at the node. The node sits 12\,meV above $E_\mathrm{F}$ because the cut misses the exact crossing by the amount given above; the nearest other bands are 0.25\,eV away and lie outside the panel.}\label{fig:dft_symmetry}
\end{figure}

Finally, we discuss the symmetry and the topology of the electronic structure of Yb$_3$Pd$_4$Bi$_8$.  As the space group $Pbam$ is centrosymmetric and the compound is non-magnetic, the product of inversion and time reversal makes every band twofold degenerate throughout the zone, which excludes Weyl nodes by symmetry. The spin--orbit calculation, in which  560 spinor states are obtained independently, find the degeneracy realised to better than 0.3\,meV over all the bands and the 348 $k$ points (Fig.~\ref{fig:dft_symmetry}a).
$Pbam$ is nonsymmorphic: The $b$- and $a$-glides carry non-primitive translations in the $ab$ plane, and their phases force additional band degeneracy on the zone boundary. Without spin--orbit coupling the whole $k_x=1/2$ and $k_y=1/2$ planes are fourfold degenerate, forming two nodal surfaces (Fig.~\ref{fig:dft_symmetry}a). Spin--orbit coupling removes the surfaces but leaves five fourfold nodal lines, X--S, S--Y, U--R, R--T and S--R: those that either lie in the $k_z=0$ or $k_z=1/2$ mirror plane or lie where the two glide planes intersect. Three fourfold branches cross the Fermi level along S--R and one along R--T (Fig.~\ref{fig:dft_symmetry}b), so that Yb$_3$Pd$_4$Bi$_8$ hosts symmetry-enforced Dirac points at $E_\mathrm{F}$. 
We note that the compound is metallic, so that a $\mathbb{Z}_2$ invariant is not defined. The parity products at $\Gamma$ and Z are both $-1$, indicating no band inversion between them. 

A nodal line is fourfold degenerate along its entire length, so a path running along the specific S--R high-symmetry path cannot by itself establish that the crossing at $E_\mathrm{F}$ is a Dirac point: the dispersion has to be examined in the plane perpendicular to the  line. We therefore computed two further cuts through the crossing nearest $E_\mathrm{F}$, one along $a^*$ and one along $b^*$ at $k_z=0.123$ (Fig.~\ref{fig:dft_symmetry}c). 
On the line, the four states are degenerate to better than 0.1\,meV, the precision to which the eigenvalues are reported; away from it they separate into two Kramers doublets whose splitting grows linearly in $|\delta k|$ over the first 0.02\,\AA$^{-1}$. The slopes give transverse velocities $\hbar v_a=0.50$ and $\hbar v_b=0.54$\,eV\,\AA, that is $0.76$ and $0.83\times10^{5}$\,m\,s$^{-1}$. These differ by only 8\%, so the
cone is nearly isotropic in the transverse plane even though the lattice parameters $|a^*|$ and $|b^*|$ are very different. These group velocities are about an order of magnitude smaller than the Dirac velocity of graphene, as expected for a crossing built from Bi $6p$ and Pd $4d$ states. The degeneracy at $E_\mathrm{F}$ is a genuine fourfold Dirac crossing, carried by a nodal line that runs along $c^*$ and disperses linearly in the two perpendicular directions. Along the line itself, the same quartet disperses with $\hbar v = 2.2$\,eV\,\AA, about four times the transverse velocities, so the cone is strongly anisotropic overall even though it is nearly isotropic in the plane perpendicular to the line. The same argument applies to the other three crossings of $E_\mathrm{F}$: the fourfold degeneracy is enforced along the whole of S--R and R--T, each crossing is isolated from the rest of the spectrum by 180 to 600\,meV, and each disperses linearly along its own line with $\hbar v$ between 2.1 and 4.6\,eV\,\AA. Only the crossing at $k_z=0.1265$ was cut perpendicular, however, so the transverse cone is verified there alone; the transverse velocities at the other three will differ, and the crossing on R--T has its transverse plane spanned by $b^*$ and $c^*$ rather than by $a^*$ and
$b^*$.

\section{Conclusion}

We have successfully grown single crystals of Yb$_3$Pd$_4$Bi$_8$. It is isostructural to Ca$_3$Pd$_4$Bi$_8$ and crystallizes in orthorhombic space group 55, $Pbam$. The magnetic response is diamagnetic, indicating that the Yb has a non-magnetic Yb$^{2+}$ configuration. The electrical resistivity is metallic and can be explained by electron-phonon scattering over the whole temperature range. These findings are supported by DFT calculations and demonstrate that the Pd--Bi atomic network of the $R_3$Pd$_4$Bi$_8$ structure type accommodates a stable divalent Yb ion while preserving its characteristic network of interconnected PdBi$_6$ trigonal prisms and metallic character. 
Electronic structure calculations with fully relativistic spin-orbit coupling show that the nonsymmorphic $Pbam$ crystal structure enforces fourfold-degenerate nodal lines on the zone boundary, four of which cross E$_F$ and give symmetry-enforced Dirac crossings.
As a non-magnetic analog, Yb$_3$Pd$_4$Bi$_8$ provides an important baseline for disentangling electronic and magnetic contributions, enabling comparison with future rare-earth members of this series. The successful stabilization of this compound thus motivates the search for Eu-based analogs, where the large localized Eu$^{2+}$ moment can be incorporated into the same Bi-rich metallic framework. Such compounds would offer a platform to explore the interplay between local-moment magnetism and strong spin–orbit coupling, which is expected to be mediated by conduction electrons via the RKKY interaction in an isostructural setting.

\section*{Acknowledgements}

We thank Matthieu Quénot for the EDS analyses and Johnathan Bulled for support at the ID28 beamline. We acknowledge financial support from the French National Research Agency (ANR) under the project KONWEY (ANR-24-CE97-0006). M.A.P. and V.T. acknowledge funding from the Laboratoire d’excellence LANEF in Grenoble (ANR-10-LABX-51-01). S.R. thanks the support from the ANR under the project SUPERNICKEL (Grant No.ANR-21-CE30-0041-04) and also the European Research Council (ERC) under the European Union's Horizon 2020 research and innovation program (Grant Agreement No. 865826). The research at UdeM received support from the Natural Sciences and Engineering Research Council of Canada (Canada), as well as the RQMP through the grant \url{https://doi.org/10.69777/309032}. The theory work was supported by the U.S. Department of Energy, Office of Basic Energy Sciences, Division of Materials Science and Engineering, grant DE-SC0022288. 



\bibliography{biblio, refs_georg, refs_sitaram, refs_dft}



\end{document}